\documentstyle[12pt,fleqn]{article}
\catcode`\@=11
\begin{document} \openup6pt

\title{{Gravitational collapse of null fluid on the brane}}

\author{Naresh Dadhich \thanks{email: nkd@iucaa.ernet.in}
\\Inter-University Center for Astronomy and Astrophysics, \\
Post Bag 4, Ganeshkhind, Pune - 411 007, INDIA
 \and and \\ S. G. Ghosh \thanks{Author to whom all correspondence should be
directed; email:
sgghosh@hotmail.com}
\\Department of Mathematics, Science College, Congress Nagar, \\
Nagpur-440 012, INDIA}

\vspace{2in}
\date{}

\maketitle
\begin{abstract}
We first obtain the analogue of Vaidya's solution on the brane for studying 
the collapse of null fluid onto a flat Minkowski cavity on the brane. Since 
the back-reaction of the bulk onto the brane is supposed to strengthen gravity 
on the brane, it would favour formation of black hole as against naked 
singularity. That is the parameter window in the initial data set giving rise 
to naked singularity in the $4D$ Vaidya case would now get partially covered.
\end{abstract}


PACS No(s): 04.50.+h;98.80.Cq;12.10.-g;11.25.Mj;04.70.Dw;04.70.Bw

\section{Introduction}
General relativity (GR) is a successful theory of gravitation in
the energy range accessible to observation. However, at very high
energies where all fields are expected to move towards
unification, considerations of modification to GR would be quite in order. 
In this
context, string theory proposes a grand framework for unification
of all forces and it requires dimensions larger than the usual four.
That is, at ultra high energy, there should be a new theory of
gravitation, of which GR should turn out to be the low energy limit. Of
course we are far from that theory, yet it has inspired
considerable interest in higher dimensional gravity.

The canonical scenario in the higher dimensional models is that
the matter fields are confined to a $1+3$-dimensional $3$-brane in $1 + 3 + d$
dimensional spacetime, while gravitational field propogates in the
extra $d$ dimensions as well \cite{rm,1rm}. Earlier the extra
dimensions were taken to be very compact, but recently this
condition has also been relaxed by Randall and Sundrum \cite{2rm}.
They have shown for $d=1$ gravitational field could be confined
in the neighbourhood of the $3$-brane without any restriction of
finiteness of the fifth extra dimension (see also \cite{3rm}).
Shiromizu, Maeda and Saski \cite{4rm} have worked out an elegant
geometric framework for the R-S brane model. The $5$-dimensional
spacetime is called bulk while the Universe we live in brane, in
which the standard models and theories of fields and matter enjoy
the observational support. In the R-S model, it is envisaged that
the bulk is a solution of the Einstein equation with negative cosmological 
constant, $\Lambda$. Gravitational field gets "reflected" from the bulk
onto the brane through the projection of the bulk Weyl curvature
on the brane. This projection would appear as trace free matter
field on the brane. By "reflection" from bulk the effective
Einstein equation on the brane would in addition to the usual
stresses include quadratic terms in them plus the trace free part
which arises from free gravitational field \cite{4rm}. For
seeking solution of a problem in the brane world model is first
to find the solution with negative $\Lambda$ in the bulk and then
solution of the effective equation on the brane which has to be
matched with the bulk solution satisfying the proper boundary
conditions \cite{4rm}.

In view of the involved process of finding solution, it would be
understandably quite difficult to find the complete solution. A
solution for a black hole on the brane has recently been proposed
\cite{dmpr}, which takes into account the trace free matter field
induced on the brane from the bulk. It is given by the
Reissner-Nordstr$ \ddot{o}$m metric for a charged black hole.
Here the induced charge is the tidal gravitational charge, and
not the electric charge, and it is the measure of "reflected"
gravitational field energy as a matter field on the brane. The
solution was obtained by solving for the trace free condition and
null energy condition on the brane. It has not been matched with
the $\Lambda$ vacuum in the bulk. The null energy condition is
required to hold good on the horizon and not necessarily
everywhere off the horizon. It is however an approximate solution
which is good in the neighbourhood of the horizon \cite{tet}. It
has been argued that induced energy density on the brane must be
negative so as to strengthen hole's gravity and as well as not to
radically change the singularity structure of spacetime. When an
isolated object is sitting in an energy distribution which
vanishes asymptotically, it would contribute in line with
object's gravity only when its energy density is negative
\cite{dad}. The modification ensuing from bulk's back-reaction
would thus strengthen black hole's gravitational field. It would
therefore be pertinent to study gravitational collapse in this
model. Would the parameter window in the initial data set giving
rise to naked singularity shrink owing to effective strengthening
of gravity? This is the question we wish to address in this letter.

The cosmic censorship conjecture (CCC) is one of the most important unresolved 
question in classical general relativity \cite{rp}. It is of course 
well-known that gravitational collapse under very general conditions in GR 
ultimately 
leads to a singularity \cite{he}. The question arises whether singularity so 
formed is 
visible to asymptotic observer or hidden behind an event horizon. The CCC 
states that the latter is the case. In its strong form, it demands spacetime 
to be globally hyperbolic (and hence its invisibility to any observer), while 
the weak form demands formation of a horizon and its invisibility only to an 
asymptotic 
observer (see \cite{r1}, for reviews on CCC). 
In view of its profound importance, the question also attains pertinence 
in situations that may call for modification of GR. We shall therefore like to 
address this question for the brane world model. That is, what effect would 
the back-reaction of the bulk cause on gravitational collapse on the brane? 

We shall consider collapse of the null fluid described by the
Vaidya solution onto a flat Minkowski cavity. The exterior to the
Vaidya region would be the Schwarzschild region. That is the
brane-generalized Vaidya would be matched to the
brane-generalized Schwarzschild solution. For this we shall first
find the analogue of the Vaidya solution on the brane taking into
account the bulk induced trace free matter field. In finding the
end product of collapse, we need only to know the generalized
Vaidya solution on the brane. In the next section, we write the
effective Einstein equation on the brane and the generalized
Vaidya solution on the brane, which would be followed by
discussion of occurrence of naked singularity. We conclude with a
discussion.

\section{Vaidya solution on the brane}
The Einstein equation for the 5-dimensional bulk spacetime with a
negative $\Lambda$ and brane energy-momentum as source
\cite{rm,4rm}:
\begin{equation}
\tilde{G}_{AB} = \tilde{\kappa}^2 [ - \tilde{\Lambda}
\tilde{g}_{AB} + \delta(\xi)({- \lambda g_{AB} + T_{AB}})]
\end{equation}
The tildes indicate the 5-dimensional bulk analogous of the
standard GR quantities, and $\tilde{\kappa}^2 = 8 \pi/
\tilde{M}_p^3$, where $\tilde{M}_p$ is the fundamental
5-dimensional Planck mass, which is much less than the brane
Planck mass, $M_p = 1.2 \times 10^{19} \; GeV$. The brane is
located at $\xi = 0$ which suggests the coordinates $(x^\mu,
\xi)$ where $x^\mu = (t, x^i)$ are the spacetime coordinates on
the brane. $\lambda$ is the brane tension and induced metric on
the brane is given by $g_{AB} = \tilde{g}_{AB} - n_An_B$ where
$n_A$ is the space-like unit normal to the brane. The brane stress
tensor $T_{AB}$ refers to the brane confined matter fields with
$T_{AB}n^A = 0$.

The effective equation on the brane is obtained by Shiromizu {\it
et al.} \cite{4rm} by using the Gauss-Codacci equations, the
matching conditions and the $Z_2$ symmetry of the spacetime. 
It reads as follows:
\begin{equation}
G_{\mu\nu} = -\Lambda g_{\mu\nu} + \kappa^2 T_{\mu\nu} +
\tilde{\kappa}^4 S_{\mu \nu} - \cal E_{\mu \nu},
\end{equation}
where
\begin{equation}
S_{\mu\nu} = \frac{1}{12}T^\alpha_\alpha T_{\mu\nu} -
\frac{1}{4}T_{\mu\alpha}T_\nu^\alpha +
\frac{1}{24}g_{\mu\nu}\left[ 3T_{\alpha\beta}T^{\alpha\beta} -
(T^\alpha_\alpha)^2 \right]
\end{equation}
and

\begin{equation}
{\cal{E}}_{AC} = \tilde{C}_{ABCD} \; n^B \; n^D.
\end{equation}
Here $\tilde{C}_{ABCD}$ is the bulk Weyl curvature and hence ${
{\cal{E}}_{AB} = {\cal{E}}_{BA} }$, ${{\cal{E}}^A_A} = 0$, ${{
\cal{E}}_{AB}}n^B   = 0$. The induced modifications from the bulk
are of two kinds, one the quadratic stress corrections and the
other nonlocal effects of the free gravitational filed in the
bulk transmitted through the projection of the bulk Weyl
curvature. Thus $\cal{E}_{\mu\nu}$ represents reflected nonlocal
gravitational degrees of freedom from bulk to the brane and it
includes tidal (Coulomb), gravito-magnetic as well as transverse
traceless (gravitational wave) effects.

The energy scales are related by
\begin{equation}
\lambda = 6\frac{\kappa^2}{\tilde{\kappa}^4}, ~\Lambda =
\frac{4\pi}{\tilde{M}_p^3} \left[ \tilde{\Lambda} + \left(
\frac{4\pi}{3\tilde{M}_p^3} \right)\lambda^2 \right]
\end{equation}
where $\kappa^2 = 8 \pi/M_p^2$.

For obtaining the analogue of the Vaidya solution \cite{pc} on the
brane, we note that $T_{\mu\nu} = \sigma k_\mu k_\nu, ~k_\mu
k^\mu = 0$ where $\sigma$ is the density of null fluid. We set
$\Lambda = 0$, and the equation we need to solve is
\begin{equation}
G_{\mu\nu} = \kappa^2 \sigma k_\mu k_\nu - \cal E_{\mu\nu}
\end{equation}
because for the null fluid $S_{\mu \nu} = 0$. The solution in
$(v,r, \theta, \phi)$ coordinates is given by
\begin{equation}
ds^2 = - (1 - \frac{2 m(v)}{r} - \frac{e^2(v)}{r^2}) dv^2 + 2 dv
dr + r^2 d \Omega^2  \label{eq:me}
\end{equation}
where $ d\Omega^2 = d \theta^2+ sin^2 \theta d \phi^2$, $v$
represents advanced Eddington time, in which $r$ is decreasing
towards the future along a ray $v=const.$, the two arbitrary
functions $m(v) = M(v)/M_p^2$ and $e(v)^2 = M^2(v)/\tilde M_p^4$ refer
respectively to the mass and reflected tidal charge at advanced
time $v$. They would be only restricted by the proper energy
conditions \cite{dad}. In contrast to the positive energy density of electric 
field for the charged black hole, the energy density of reflected 
gravitational field is negative which accounts for the negative sign before 
$e^2/r^2$ term in the metric.

We have thus obtained an analogue of the Vaidya solution on the
brane in the same manner as the Schwarzschild's analogue
\cite{dmpr}. It would also attract the same criticism that it
would be valid only close to the horizon. In our analysis the behaviour of 
solution near the horizon is most relevant. Thus even though this solution 
may not be valid globally, it would be good enough for our purpose here to
probe occurrence of naked singularity or black hole.

\section{Collapse on the brane}

In this section, we employ the above metric for investigation of
formation of black hole or naked singularity in collapse of null
fluid in the context of CCC. The physical situation is that of a
radial influx of null fluid in an initially empty region of the
Minkowski spacetime. The first shell arrives at $r=0$ at time
$v=0$ and the final at $v=T$. A central singularity of growing
mass is developed at $r=0$.
  For $ v < 0$ we have $m(v)\;=\;e(v)\;=\;0$, i.e., an empty
 Minkowski metric, and for $ v > T$, $\dot{m}(v)\;=\;\dot{e}(v)\;=\;0$, $m(v)\; and
\;e^2(v) $ are positive definite.  The metric for $v=0$ to $v=T$
is the brane-generalised Vaidya, and for $v>T$ we have the
brane-generalised Schwarzschild solution.

In order to proceed further we would require the specific forms of
the functions $m(v)$ and $e^2(v)$ which we choose as follows:
\begin{equation}
m(v) = \left\{ \begin{array}{ll}
        0,                      &       \mbox{$ v < 0$}, \\
        \lambda v (\lambda>0)   &       \mbox{$0 \leq v \leq T$}, \\
        m_{0}(>0)               &       \mbox{$v >  T$}.
                \end{array}
        \right.                         \label{eq:mv}
\end{equation}
and
\begin{equation}
e^2(v) = \left\{ \begin{array}{ll}
        0,                      &       \mbox{$v < 0$}, \\
        \mu^2 v^2 (\mu^2>0)     &       \mbox{$0 \leq v \leq T$}, \\
        e_{0}^2 (>0)            &       \mbox{$v >  T$}.
                \end{array}
        \right.                         \label{eq:ev}
\end{equation}
Then the space-time is self-similar \cite{ss}, admitting a
homothetic Killing vector
\begin{equation}
\xi^a = r \frac{\partial}{\partial r}+v \frac{\partial}{\partial
v} \label {eq:kl1}
\end{equation}
which is given by the Lie derivative
\begin{equation}
\L_{\xi}g_{ab} =\xi_{a;b}+\xi_{b;a} = 2 g_{ab}
\end{equation}
$\L$ denotes the Lie derivative. Let $K^{a} = dx^a/dk$ be the
tangent vector to the null geodesics, where $k$ is an affine
parameter. Then
\begin{equation}
g_{ab}K^a K^b=0.
\end{equation}
It follows that along null geodesics, we have
\begin{equation}
\xi^a K_{a} = r K_r + v K_v = C \label{eq:kl2}
\end{equation}

\noindent Following \cite{dj}, we introduce
\begin{equation}
K^v = \frac{P}{r}       \label{eq:kv}
\end{equation}
and, from the null condition, we obtain
\begin{equation}
K^r = \left[ 1 -  \frac{2 m(v)}{r} - \frac{e^2(v)}{r^2} \right]
\frac{P}{2r}.    \label{eq:kr}
\end{equation}

To study the singularity we employ the method developed by
Dwivedi and Joshi \cite{dj}.  Consider the equation for  radial,
outgoing null geodesics
\begin{equation}
\frac{dr}{dv} = \frac{1}{2} \left[1 -  \frac{2 m(v)}{r} -
\frac{e^2(v)}{r^2}
 \right].                \label{eq:de1}
\end{equation}
This is an ordinary differential equation with a singular point
$v=0, \; r=0$.  This singularity is (at least locally) naked if
there are geodesics starting at it with a definite
tangent.  If no geodesic exists, then singularity is not naked
and strong CCC holds. To investigate the behaviour near singular
point, define
\begin{equation}
y = \frac{v}{r}.  \label{eq:sv}
\end{equation}
Eq. (\ref{eq:de1}), upon using  eqs. (\ref{eq:mv}), (\ref{eq:ev})
and (\ref{eq:sv}), turns out to be
\begin{equation}
\frac{dr}{dv} = \frac{1}{2} \left[1 - 2 \lambda y - \mu^2 y^2
\right].         \label{eq:de2}
\end{equation}
If singularity is naked, there exists some value of $\lambda$ and
$\mu$ such that at least one positive finite value $y_0$ exists which
solves the algebraic equation
\begin{equation}
y_{0} = \lim_{r \rightarrow 0 \; v\rightarrow 0} y =
\lim_{r\rightarrow 0 \; v\rightarrow 0} \frac{v}{r}. \label{eq:lm1}
\end{equation}
Using (\ref{eq:de2}) and L'H\^{o}pital's rule we get
\begin{equation}
y_{0} = \lim_{r\rightarrow 0 \; v\rightarrow 0} y =
\lim_{r\rightarrow 0 \; v\rightarrow 0} \frac{v}{r}=
\lim_{r\rightarrow 0 \; v\rightarrow 0} \frac{dv}{dr} =
\frac{2}{1 - 2 \lambda y_{0} - \mu^2 y_{0}^2}    \label{eq:lm2}
\end{equation}
which can be written in explicit form as,
\begin{equation}
 \mu^2 y_{0}^3 + 2 \lambda y_{0}^2 - y_{0} + 2 = 0.
\label{eq:ae}
\end{equation}
The central shell focusing singularity is naked, if eq. (\ref{eq:ae})
 admits one or more positive real roots.  Hence in the absence
of positive real roots, the collapse will always lead to a black
hole. It can be shown that real roots of eq. (\ref{eq:ae}) exist
only when $\Delta = \lambda^2 - 18 \lambda \mu^2 - 16 \lambda^3 +
\mu^2 - 27 \mu^4 \geq 0$. It is easy
to check that two of three real roots of eq. (\ref{eq:ae}) are
always positive. We have summarised the results in the following
table. Hence we can say that the nature of the singularity (naked
or covered) depends on the behaviour of $\Delta$. The behaviour of
$\Delta$ in $(\lambda, \mu)$ space is shown in Figs. I-III.

\vspace{.4in}

\noindent {\bf Table I: Roots $X_0$ for different $\lambda$ and
$\mu$}

\begin{tabular}{|c|l|l|} \hline
$\lambda$ & $ \mu$ & $Roots \; (X_0)$ \\ \hline $0.06$ &
$0.03$&$3.57311, \; 4.40124$ \\ \hline $0.05$ & $0.05$ &
$2.90554,\;
5.66840$ \\ \hline $1/16$ & $0.00$ & $4, \; 4$ \\
\hline $0.0005$ & $0.19$ & $2.80212, \; 3.25127$               \\
\hline $0.0$  & $0.10$ &   $2.09149, \; 8.78885$               \\
\hline

\end{tabular}

\vspace{.2in}

It can be seen that collapse always leads to naked singularity if
$\lambda \leq 0.06$ and $\mu \leq 0.03$ (Fig. I). Whereas for
$\mu > 0.19$ it is always a black hole for all values of
$\lambda$ (Fig. II).  When $\mu=0$, we come back to the Vaidya metric, and eq.
 (\ref{eq:ae}) admit positive roots for $0 < \lambda \leq 1/16$
and hence singularities  are naked for this range of $\lambda$ \cite{r1}.

\input{epsf}
\begin{figure}
\epsffile{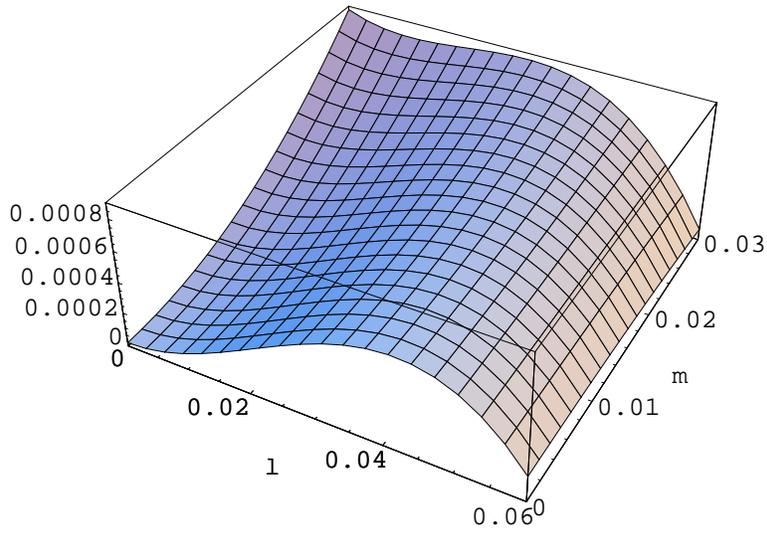}
\caption{NS at all points of this $(\lambda \leq .06,
\mu \leq .03)$ space}
\end{figure}
\input{epsf}
\begin{figure}
\epsffile{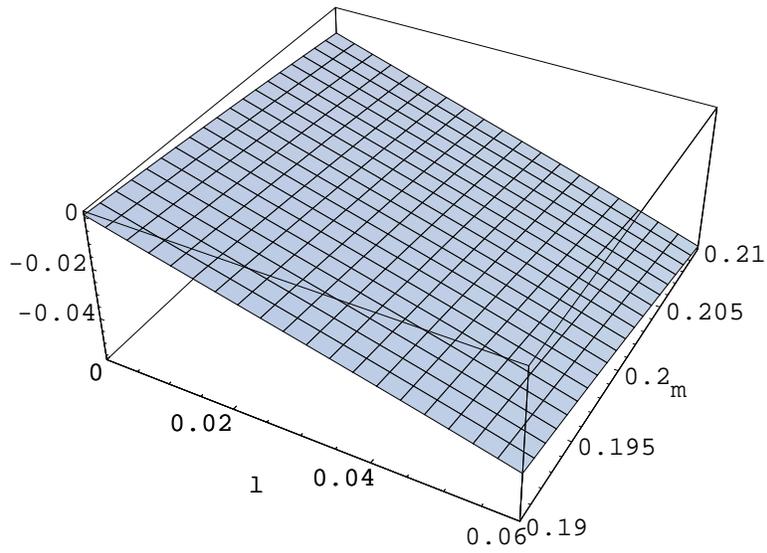}
\caption{BH occurrs at all points of the $(\lambda,
\mu)$ space for which  $\mu > .19$}
\end{figure}
\input{epsf}
\begin{figure}
\input{epsf}
\epsffile{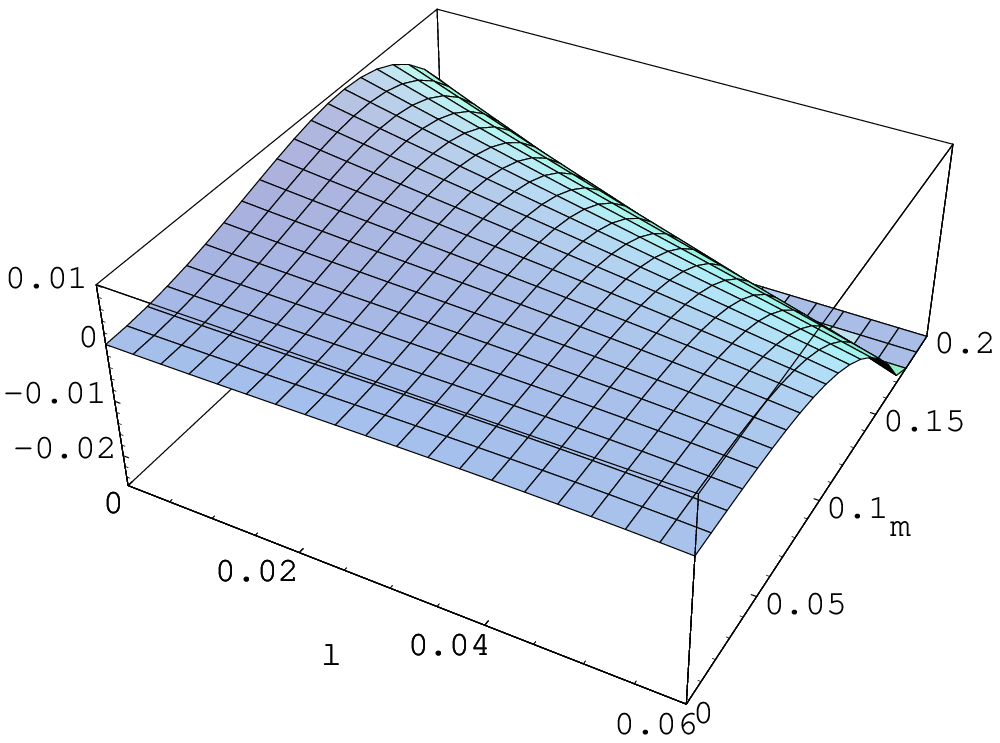}
\caption{Both NS and BH occur}
\end{figure}

\paragraph{Strength of Naked Singularities:}
A singularity is termed gravitationally strong or simply strong,
if it destroys by crushing or stretching any object to zero volume
which falls into it. A sufficient condition \cite{ck} for a
strong singularity as given by Tipler \cite{ft} is that for at
least one non-spacelike geodesic with affine parameter $k$, in
limiting approach to singularity, we must have
\begin{equation}
\lim_{k\rightarrow 0}k^2 \psi = \lim_{k\rightarrow 0}k^2 R_{ab}
K^{a}K^{b} > 0 \label{eq:sc}
\end{equation}
where $R_{ab}$ is the Ricci tensor. Eq. (\ref{eq:sc}) can be
expressed as
\begin{equation}
\lim_{k\rightarrow 0}k^2 \psi = \lim_{k\rightarrow 0} 2
\left[\dot{m}(v) + \frac{e(v) \dot{e}(v)}{r} \right] \left[
\frac{kP}{r^2} \right]^2 \label{eq:sc1}
\end{equation}
Our purpose here is to investigate the above condition along
future directed null geodesics coming out from the singularity.
For this, solution $P$ is required. Eq. (\ref{eq:kl2}), because
of eqs. (\ref{eq:mv}), (\ref{eq:kv}) and (\ref{eq:kr}), yields
\begin{equation}
P = \frac{2 C}{2- y +2 \lambda y^2 + \mu^2 y^3} \label{eq:ps}
\end{equation}
 and the geodesics are then completely
known. Further, we note that
\begin{equation}
\frac{dy}{dk} = \frac{1}{r} K^v - \frac{X}{r} K^r \label{eq:xk}
\end{equation}
which, on inserting the expressions for $K^v$ and $K^v$, become
\begin{equation}
\frac{dy}{dk} = (2 - y + \lambda y^2 + \mu y^3) \frac{P}{2r^2} =
\frac{C}{r^2}.
 \label{eq:xk1}
\end{equation}
Using the fact that as singularity is approached, $k \rightarrow
0$, $r \rightarrow 0$ and  $X \rightarrow a_{+}$ (a root of
(\ref{eq:ae})) and using  L'H\^{o}pital's rule, we observe
\begin{equation}
\lim_{k\rightarrow 0} \frac{kP}{r^2} = \frac{2}{1+\mu y_{0}^2}
\end{equation}
and hence eq. (\ref{eq:sc1}) gives
\begin{equation}
\lim_{k\rightarrow 0}k^2 \psi = \frac{8 \lambda }{(1+\mu
y_{0}^2)^2} >0.
\end{equation}
Thus along radial null geodesics strong curvature condition is
satisfied. Therefore, one may say that generically, the naked
singularity is gravitationally strong 
\cite{ft}. Having seen that the naked singularity in our model
is a strong curvature singularity, we also examine its scalar
polynomial character. The Kretschmann scalar  ($K = R_{abcd}
R^{abcd}$, $R_{abcd}$ is the Riemann tensor) with the help of
eqs. (\ref{eq:mv}) and (\ref{eq:ev}), takes the form
\begin{equation}
K = \frac{48}{r^4} \left[ \lambda^2 y^2 + 2 \lambda \mu^2 y^3 +
\frac{7}{6} \mu^4 y^4 \right] \label{eq:ks1}
\end{equation}
which diverges at the naked singularity and hence the singularity
is a scalar polynomial singularity \cite{he}.

Finally, we compute the Ricci scalar ($R = R_{a}{}^{a}$)
 for the metric (\ref{eq:me}) is
\begin{equation}
R = 4 \left[ \frac{\mu^4 y^4}{r^4} \right]
\label{eq:rs1}
\end{equation}
The Ricci scalar, which vanishes in $4D$ Vaidya case, has the same divergence 
rate  as the Kretschmann scalar. 

\section{Discussion}

 We have generalized the Vaidya solution in the same spirit as
the Schwarzschild solution done in \cite{dmpr} to describe a null
fluid on the brane. The reflected energy density of free
gravitational field from the bulk is negative and hence it would
contribute positively to gravity of the collapsing null radiation
\cite{dad}. This should cover part of the parameter window in the
initial data set for naked singularity. This is what has been demonstrated.gv 
That is, the parameter set which gave rise to naked singularity in
GR collapse may now lead to black hole on the brane. The window
gets covered with the increased strength of the reflected tidal
charge. There exists a threshold value for $\mu$, as shown in
Fig. II, the parameter window gets fully covered ensuring formation of black 
hole for all values of $\lambda$. 
That is when $\mu>0.19$ the CCC is always respected. On the other hand the 
CCC is always violated for $\lambda \leq0.06$ and $\mu\leq0.03$ (Fig. I).  
The singularity is always Tipler strong curvature singularity. Of course 
the tidal charge goes as $M^2/ \tilde{M}_p^2$ which cannot take arbitrarily 
large value. The important point is that collapse on the brane would favour 
black hole in comparison to naked singularity. This is what was however 
expected because of strengthening of gravity resulting from the back-reaction 
of the bulk. 

 One of the important issues about existence of naked singularities is their 
stability, and there exists no well formulated criterion for it. In such a 
situation, it is prudent to examine the question under various conditions. For 
instance very recently, the effect of immersing the Vaidya null fluid 
collapsing sysetm into a constant potential bath was considered \cite{jdj} 
and was shown that naked singularity does persist. The present consideration 
could as well be considered from this viewpoint. That is, how does it fare 
with respect to the possible high energy modifications to GR in the brane 
world model? It seems to be stable with partially covered parameter window in 
the initial data set for naked singularity.  
 
 It would thus be fair to say that the brane inspired modification of GR at 
high energies seems to have tendency towards the CCC. However, the parameter 
window allowing violation of the CCC gets fully covered only for very large 
``reflected'' charge, which may not be realistically sustainable. This is 
the issue for further probing in future. \\ 

\noindent
{\bf Acknowledgement:} One of the authors (SGG) would
like to thank IUCAA for hospitality and Science College, Congress
Nagar, Nagpur for granting leave.

\noindent


\begin{thebibliography}{99}
\bibitem{rm} R. Maartens, hep-th/0004166
\bibitem{1rm} J. Polchinski, {\it Phys. Rev. Lett.} {\bf
75} (1995) 4724; \\ P. Horava and E. Witten, {\it Nucl. Phys.} {\bf
B460} (1996) 506.
\bibitem{2rm} L. Randall and R. Sundram, {\it Phys. Rev. Lett.} {\bf
83} (1999) 4690.
\bibitem{3rm}K. Akama, hep-th/0001113; \\
 V. A. Rubakov and M. E. Shaposhnikov, {\it Phys. Lett.} {\bf
B125} (1983) 139; \\ M. Viser, {\it Phys. Lett.} {\bf B159} 
(1985) 22; \\ M. Gogberashvili, {\it Europhys. Lett.} {\bf 49} 
(2000) 369.
\bibitem{4rm} T. Shiromizu, K. Maeda, and M. Sasaki, 
{\it Phys. Rev.} {\bf D 62} (2000) 024012.
\bibitem{dmpr} N. Dadhich, R. Maartens, P. Papadopoulos and V. Rezania,
{\it Phys. Lett.} {\bf B487} (2000) 1
\bibitem{tet} A. Chamblin, H. S. Reall, H. Shankai and
T. Shiromizu, hep-th/0008177.
\bibitem{dad} N. Dadhich, {\it Phys. Lett.} {\bf B492} (2000) 357
\bibitem{rp} R. Penrose, {\it Riv. del Nuovo Cim.} {\bf 1}, 252
(1969); \\ {\it ibid.} in: S. W. Hawking and W. Israel (Eds.), 
{\it General Relativity, an Einstein Centenary Volume}, 
Cambridge University Press, England, 1979.
\bibitem{he} S.W. Hawking and G.F.R. Ellis, {\it The Large
Scale Structure of Spacetime}, Cambridge University Press,
Cambridge, 1973.
\bibitem{r1} P.S. Joshi, {\it Global Aspects in Gravitation and
Cosmology}, Clendron Press, Oxford, 1993; \\
 C.J.S. Clarke, {\it Class. Quantum Grav.} {\bf 10} 
 (1993) 1375; \\ R.M. Wald, gr-qc/9710068 (1997); \\ T.P. Singh, {\it J.
Astrophys. Astr.} {\bf 20} (1999) 221; \\
 S. Jhingan  and G. Magli, gr-qc/9903103 (1999); \\
P. S. Joshi, gr-qc/0006101 (2000).
\bibitem{pc} P. C. Vaidya, {\it Proc. Indian Acad. Sci.} {\bf
A33} (1951) 264; \\ Reprinted in {\it Gen. Rel. Grav.} {\bf 31} 
(1999) 119.
\bibitem{ss} A spherical symmetric space-time is self similar if
$ g_{tt} (ct, cr) = g_{tt}(t,r)$ and  $ g_{rr} (ct, cr) =
g_{rr}(t,r)$ for every $c>0$.  A self similar spacetime is
characterized by the existence of a homothetic Killing vector.
\bibitem{dj} I. H. Dwivedi and P. S. Joshi, {\it Class. Quantum
Grav.} {\bf 6} (1989) 1599.
\bibitem{ck} C. J. S. Clarke and A. Kr\'{o}lak, {\it J. Geom.
Phys.} {\bf 2} (1986) 127.
\bibitem{ft} F. J. Tipler, C. J. S. Clarke, and G. F. R. Ellis,
in:  A. Held (Ed.), {\it General Relativity and Gravitation}, Plenum,
New York, 1980.
\bibitem{jdj} S. Jhingan, N. Dadhich, and P. S. Joshi,
 gr-qc/0010111 - to appear in Phys. Rev. D.

 
\end{thebibliography}
\end{document}